\documentclass[12pt]{elsarticle}

\usepackage{amssymb}
\usepackage[table,xcdraw]{xcolor}
\usepackage{graphicx}
\usepackage{booktabs}
\usepackage{array}
\usepackage{amsmath}
\usepackage[caption = false]{subfig}
\usepackage{multirow}
\usepackage[normalem]{ulem}
\useunder{\uline}{\ul}{}

\journal{Artificial Intelligence in Medicine}

\begin{document}

\begin{frontmatter}



\title{Uncertainty-Aware
Temporal Self-Learning (UATS): Semi-Supervised Learning for Segmentation of Prostate Zones and Beyond}

\author{Anneke Meyer\corref{cor1}\fnref{label1}} 
\author{Suhita Ghosh\corref{cor1}\fnref{label1}}

\author{Daniel Schindele\fnref{label2}}
\author{Martin Schostak\fnref{label2}} 
\author{Sebastian Stober\fnref{label1}}
\author{Christian Hansen\fnref{label1}}
\author{Marko Rak\fnref{label1}}

\fntext[label1]{Faculty of Computer Science and Research Campus STIMULATE, University of Magdeburg, Germany}
\fntext[label2]{Clinic of Urology and Pediatric Urology, University Hospital Magdeburg, Germany}
\cortext[cor1]{Anneke Meyer and Suhita Ghosh contributed equally.}
\address{}
\nonumnote{\copyright\,2021. This manuscript version is made available under the CC-BY-NC-ND 4.0 license http://creativecommons.org/licenses/by-nc-nd/4.0/.\\}

\begin{abstract}
Various convolutional neural network (CNN) based concepts have been introduced for the prostate's automatic segmentation and its coarse subdivision into transition zone (TZ) and peripheral zone (PZ). However, when targeting a fine-grained segmentation of TZ, PZ, distal prostatic urethra (DPU) and the anterior fibromuscular stroma (AFS), the task becomes more challenging and has not yet been solved at the level of human performance. One reason might be the insufficient amount of labeled data for supervised training. Therefore, we propose to apply a semi-supervised learning (SSL) technique named uncertainty-aware
temporal self-learning (UATS) to overcome the expensive and time-consuming manual ground truth labeling. We combine the SSL techniques temporal ensembling and uncertainty-guided self-learning to benefit from unlabeled images, which are often readily available. Our method significantly outperforms the supervised baseline and obtained a Dice coefficient (DC) of up to 78.9\% , 87.3\%, 75.3\%, 50.6\% for TZ, PZ, DPU and AFS, respectively.  The obtained results are in the range of human inter-rater performance for all structures. Moreover, we investigate the method's robustness against noise and demonstrate the generalization capability for varying ratios of labeled data and on other challenging tasks, namely the hippocampus and skin lesion segmentation. UATS achieved superiority segmentation quality compared to the supervised baseline, particularly for minimal amounts of labeled data.
\end{abstract}

\begin{keyword}
Semi-Supervised Deep Learning \sep Biomedical Segmentation \sep Prostate Zones




\end{keyword}

\end{frontmatter}


\section{Introduction}

Multiparametric MRI (mpMRI) has increased the capabilities for detecting and staging prostate cancer (PCa) over the last years.  MpMRI combines functional and physiological assessment with anatomical T2-weighted (T2w) sequences. In the latter, the prostate and its interior structure can be best distinguished from other tissues. According to \cite{mcneal1981zonal}, the prostate  consists  of four anatomical zones: transition zone~(TZ), peripheral zone~(PZ), central zone (CZ) and anterior fibromuscular stroma~(AFS). 

Prostate Imaging - Reporting and Data System version~2 (PI-RADS v2) \cite{weinreb2016pi} was introduced for a more standardized acquisition and assessment of mpMRI. PI-RADS v2 includes a so-called sector map for a more standardized lesion location assignment. The sector map consists of 41 sectors of which 38 are related to prostate zones. The remaining three sectors are the seminal vesicles and the external urethral sphincter. The sector map should support standardized reporting and “facilitate precise localization for MR-targeted prostate biopsy and therapy, pathological correlation, and research” \cite{weinreb2016pi}. Furthermore, it could also provide a “roadmap for surgical dissection at the time of radical prostatectomy” \cite{weinreb2016pi}. However, having only one fixed atlas (the sector map) for the prostate that does not reflect the enlargements and different shapes of the prostate in real patients, limits its applicability. Radiologists have to transfer the lesion location of the current case to the one in the sector map \cite{greer2018all}. Consequently,  the sector map has not been found effective for the standardized communication of lesion locations, as variability among radiologists was shown to be high \cite{greer2018all}. Therefore, an automatic segmentation of the zones would be a step towards a patient-individual sector map and could increase the repeatability and consistency of reporting, having an impact on all the above mentioned applications. 

Moreover, the zonal information could be included into automatic PCa detection algorithms which has been demonstrated to increase the accuracy of the methods \cite{mehrtash2017classification,deVente.2020}. Finally, the volume of PZ and TZ can contribute to an automatic detection of benign prostatic hyperplasia (BPH).

\begin{figure}[t!]
\centering
\includegraphics[width=1\textwidth]{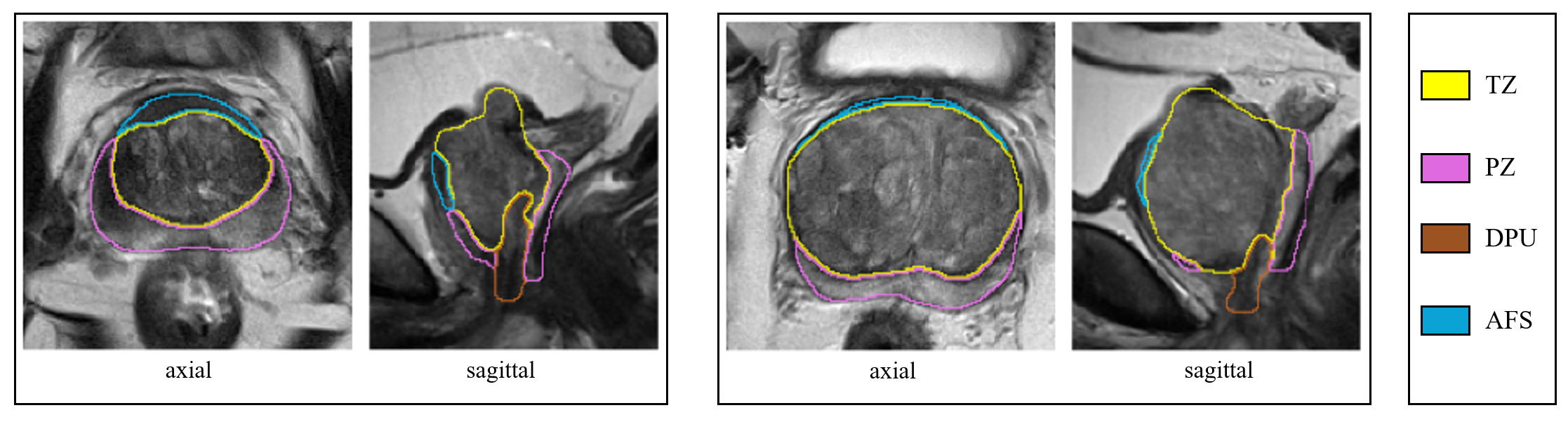}
\caption{Two cases for prostate zone segmentation in axial and sagittal view. The sagittal T2w scan was included and segmentation masks were upsampled for better visualization.} \label{fig:zones}
\end{figure}

Automatic segmentation of prostate zones is usually constrained to its coarse division into PZ and TZ.  Only our preceding work \cite{meyer2019towards} goes beyond that and proposed to automatically segment the PZ, TZ, AFS and distal prostatic urethra (DPU) in T2w MRI (see Fig.\,\ref{fig:zones}).  The DPU is needed for therapy planning, e.g., for reducing the risk of incontinence. The AFS has so far not been considered for lesion assessment, but the update of PI-RADS (Version 2.1 \cite{Turkbey.2019}) recommends to also pay attention to abnormalities in this zone. 

In our recent work, we compared our zone segmentation results to those of three physicians and found that our approach did not reach the quality of professional readers for all structures. Our objective is to close this gap, yielding expert level segmentations for all the named zones. 

The success of supervised deep learning approaches depends heavily on the amount of labeled examples used in training. The AFS zone portrays a wide variety in shape and structure (as depicted in Fig.\,\ref{fig:zones}) compared to the other zones, which indicates the need for further annotated data to improve the segmentation. Unfortunately, it is very challenging and expensive to get a considerable amount of good quality annotated data in the medical domain, as the annotations can not be obtained by crowd-sourcing and need a medical expert’s involvement. The dearth of good quality annotated data motivated exploring techniques that require limited supervision \cite{Tajbakhsh.2020}, such as weakly supervised methods \citep{xu2014weakly}, transfer learning \citep{chen2017transfer}, domain adaptation \citep{perone2019unsupervised} and semi-supervised learning (SSL) methods \cite{ouali2020overview}. Encouraged by SSL's promising results and the availability of additional unlabeled data, we set out to extend our supervised baseline by a semi-supervised method to increase segmentation quality. 

The structure of this paper is as follows. The first section gives an overview of the related work including automatic prostate zone segmentation and semi-supervised learning (Section \ref{sec:rel_work}). Section \ref{sec:methods} outlines our proposed UATS method and provides details of the used datasets. We report and discuss the results of our extensive evaluation in Section \ref{sec:evaluation} for which we carried out an ablation study and investigated the method's robustness against noise and its generality on other biomedical datasets. Lastly, we draw our conclusions and give an outlook for future work in Section \ref{sec:conclusion}.

\section{Related Work}
\label{sec:rel_work}

\begin{table}[b!]
\caption{Overview of the performance (Dice coefficient in \%) of different approaches for zonal prostate segmentation in the literature. The size of the training and test dataset is included when information was provided in the paper.}
\label{tab:SOTA}
\resizebox{\textwidth}{!}{%
\begin{tabular}{l|l|l|l|l|l|l|l|l}
\hline
\hline
\textbf{Method}               & \textbf{Year} & \textbf{Input} & \textbf{\begin{tabular}[c]{@{}l@{}}Training\\ Cases\end{tabular}} & \textbf{\begin{tabular}[c]{@{}l@{}}Test\\ Cases\end{tabular}} & \textbf{PZ} & \textbf{TZ} & \textbf{DPU} & \textbf{AFS} \\
\hline
{\ul \textbf{Semi-Automatic}} &               &                &                                                                   &                                                               &             &             &              &              \\

Makni \textit{et al.} \cite{makni2011zonal}                        & 2011          & mpMRI          &                                                                   & 31                                                            & 80.0        & 89.0        & -            & -            \\
Litjens  \textit{et al.} \cite{litjens2012pattern}                     & 2012          & mpMRI          &                                                                   & 48                                                            & 75.0 $\pm$ 7.0  & 89.0 $\pm$ 3.0  & -            & -            \\
Qiu \textit{et al.}  \cite{qiu2014dual}                        & 2014          & T2w            &                                                                   & 43                                                            & 69.1 $\pm$ 6.9  & 82.2 $\pm$ 3.0  & -            & -            \\
                              &               &                &                                                                   &                                                               &             &             &              &              \\
{\ul \textbf{Automatic}}      &               &                &                                                                   &                                                               &             &             &              &              \\
Chilali   \textit{et al.} \cite{chilali2016gland}                    & 2016          & T2w            &                                                                   & 22                                                            & 62.0 $\pm$ 7.3  & 70.2 $\pm$ 12.1 & -            & -            \\
Clark  \textit{et al.}     \cite{clark2017fully}                   & 2017          & mpMRI          & 78                                                                & 26                                                            & -           & 84.7        & -            & -            \\
Mooj \textit{et al.}   \cite{mooij2018automatic}                       & 2018          & T2w            & 53                                                                &                                                               & 60.0        & 85.0        & -            & -            \\
Meyer  \textit{et al.} \cite{meyer2019towards} & 2019          & T2w            & 78                                                                & 20                                                            & 79.8 $\pm$ 5.1  & 87.6 $\pm$ 6.6  & 75.2 $\pm$ 0.34  & 41.1 $\pm$ 14.4  \\
Rundo    \textit{et al.}   \cite{Rundo2019}                   & 2019          & T2w            & 40                                                                &                                                               & 86.6 $\pm$ 1.9  & 86.8 $\pm$ 2.1  & -            & -            \\
Zabihollahy   \textit{et al.}  \cite{Zabihollahy.2019}               & 2019          & ADC            & 100                                                               & 125                                                           & 86.1 $\pm$ 9.6  & 89.9 $\pm$ 10.7 & -            & -            \\
Zabihollahy  \textit{et al.}  \cite{Zabihollahy.2019}                 & 2019          & T2w            & 100                                                               & 125                                                           & 86.8 $\pm$ 3.7  & 93.8 $\pm$ 8.9  & -            & -            \\
Liu  \textit{et al.}     \cite{Liu.2019}                      & 2019          & T2w            & 250                                                               & 63                                                            & 74.0 $\pm$ 8.0  & 86.0 $\pm$ 7.0  & -            & -            \\
Liu     \textit{et al.}    \cite{liu2020}                    & 2020          & T2w            & 259                                                               & 45                                                            & 80.0 $\pm$ 5.0  & 89.0 $\pm$ 4.0  & -            & -            \\
Aldoj    \textit{et al.}  \cite{Aldoj2020}                     & 2020          & T2w            & 141                                                               & 47                                                            & 78.1 $\pm$ 2.5  & 89.5 $\pm$ 2.2  & -            & -           \\
\bottomrule
\end{tabular}%
}
\end{table}

In this section we outline the state-of-the-art for automatic prostate zone segmentation and semi-supervised techniques in biomedical deep learning. We conclude with a summary of our work's contributions.
\subsection{Prostate Zone Segmentation}
\noindent Prior to the rise of deep learning, prostate and prostate zone segmentation has been addressed with methods using deformable models \cite{martin2010automated}, active appearance models \cite{toth2013simultaneous,qiu2014dual}, level sets \cite{toth2013simultaneous}, atlases \cite{litjens2012pattern, chilali2016gland} or machine learning \cite{makni2011zonal, litjens2012pattern}. Initially, mpMRI was used in several approaches  \cite{litjens2012pattern, makni2011zonal}, while later methods shifted to  the use of only T2w MRI \cite{chilali2016gland,qiu2014dual,toth2013simultaneous}. 

Following the success of the 2D \cite{ronneberger2015u} and 3D U-Net  \cite{cciccek20163d}, supervised deep learning has been proposed to segment the prostate as a whole structure (e.g. \cite{Brosch.2018,Jia.2019,meyer2018auto}) and its coarse division into TZ and PZ. Most works employ variants or extensions of the U-Net. CNN-based segmentation of the zones was also performed on either T2w MRI \cite{Liu.2019, Aldoj2020, liu2020, Rundo2019} or on combinations of T2w and other MRI sequences \cite{clark2017fully, Zabihollahy.2019}. However, as mentioned above, these approaches only target the coarse division into PZ and TZ and do not consider other interior structures as AFS and DPU. Thus, it is desired to improve the segmentation quality of the supervised four-zone segmentation approach described in our previous work \cite{meyer2019towards}.
The reported performance of the prostate zone segmentation approaches presented in this section is summarized in Tab. \ref{tab:SOTA}. Because the methods were developed and evaluated on different datasets, a direct performance comparison between the methods is impossible.

\subsection{Semi Supervised Learning for Segmentation}

\begin{figure}[t]
 \centering
  \includegraphics[width=0.5\textwidth]{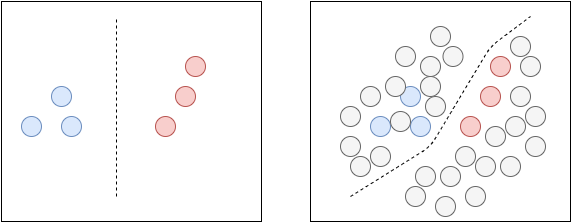}
\caption{ Illustration of the modification in the hypothesis or the decision boundary under the influence of unlabeled data. The left image portrays the decision boundary when using only labeled data, where blue circles belong to class A and red to class B. The right image portrays the change in decision boundary under the influence of unlabeled data denoted by gray circles.}
 \label{fig:ssl_basic}
\end{figure}
\noindent We base our method on SSL and leverage unlabeled data because it is often available in clinical practice. Semi-supervised methods utilize the unlabeled data to modify or re-prioritize the hypotheses obtained solely from the labeled data (see Fig.~\ref{fig:ssl_basic}). SSL techniques are based on strong assumptions that relate the properties of the data distribution with the decision function properties. Each SSL technique exploits the unlabeled data differently, governed by the based assumption. To this end, we combine two state-of-the-art SSL techniques for segmentation: self-learning and self-ensembling. To the best of our knowledge, SSL has not yet been applied to the problem of prostate zone segmentation.

\textit{Self-training} is one of the most straightforward SSL approaches and can be used with any classifier. The core mechanism is to use a trained model (initially trained on labeled data $T_L=(X_L, Y_L)$) to predict pseudo labels $Y_U$ for unlabeled data $X_U$. 
This way the labeled set for training ($T_L \cup T_U$) is enlarged, where $T_L=(X_L, Y_L)$ and $T_U=(X_U, Y_U)$. It was introduced into the biomedical domain for the segmentation of heart chambers \cite{Bai.2017}. The method works on the assumption that the model's confident predictions are correct, which complies with the cluster assumption \cite{Cheplygina.4172018}, where the instances lying in the same cluster are likely to have similar output prediction (see Fig.\,\ref{fig:ssl_basic}). 
Self-learning lacks the mechanism to rectify early mistakes, causing noisy predictions being fed into training and consequently hurting the performance. Therefore, attempts exist that examine the pseudo labels' confidence  as in \cite{Nie.2018}, which utilized a discriminator network for this purpose.

\textit{Self-ensembling} leverages the unlabelled data $X_U$ to induce regularization by enforcing consistency between different predictions, subjected to perturbations of the input data and model. %Therefore, as per the ``{no free lunch theorem}'', the success of the technique depends on whether the assumptions fit the problem structure. 
It encompasses the state-of-the-art semi-supervised methods for classification and segmentation tasks across domains \cite{oliver2018realistic}. Self-ensembling methods utilize manifold regularization, which complies with the smoothness assumption \citep{belkin2006manifold}, where it is assumed that examples lying in a similar structure (such as a manifold) are likely to belong to the same class. One example of self-ensembling is the $\pi$ model \cite{laine2016temporal}, which has been adapted for biomedical segmentation in \cite{Li.2282019}.
It penalizes the deviation between two predictions under stochastic perturbations of the input and dropout. Another example is \textit{temporal ensembling} \cite{laine2016temporal}, where regularization is derived from penalizing the current prediction's dissimilarity with an ensemble of historical predictions from different epochs under different regularization and augmentation conditions. %It has not been applied to biomedical segmentation yet. 

An extension to temporal ensembling is the two-model mean teacher \cite{Tarvainen.2017}, where the ensemble predictions are obtained from the teacher model whose weights are obtained as averaged weights of the student model. It has been adapted for biomedical segmentation \cite{perone2019unsupervised,Cui.2019,yu2019uncertainty, Li.2282019}. However, it requires two models that make the training resource-intensive for 3D data, requiring patch-based \cite{Cui.2019} or sliding window \cite{yu2019uncertainty} strategies that further limit the receptive field of the neural network to the patch size. For this reason, we decided to use the temporal ensembling strategy in our setting where the historical ensembling of predictions can be stored offline. In contrast to the $\pi$ model and the mean teacher approach, temporal ensembling has not been applied to biomedical segmentation yet.

To be clear, we will use the term self-ensembling in the following as a generic term for methods integrating the consistency loss, e.g. pi-model, temporal ensembling and mean teacher.

\subsection{Contributions}

\noindent Motivated by the potential that SSL has demonstrated in the past for various medical imaging problems, we developed a novel method named \textit{uncertainty aware temporal self-learning} (UATS), combining uncertainty-guided self-learning and concepts from temporal ensembling. The main contributions of our work are:
\begin{itemize}
    \item We propose a semi-supervised learning approach for CNN-based segmentation that leverages frequently available unlabeled data. 
    \item In our method, we combine the two SSL concepts of self-learning and temporal ensembling. To the best of our knowledge, the combination of these techniques has not been investigated yet.
    \item We show that UATS improves zonal segmentation over supervised training and outperforms state-of-the-art SSL techniques.
%    \item We are the first to implement temporal ensembling for the task of segmentation.
    \item  We demonstrate UATS' potential to improve upon supervised baselines on two other important biomedical datasets for different amounts of labeled data. 
    \item  We follow the spirit of open science, making our source code available on GitHub\footnote{https://github.com/suhitaghosh10/UATS}.
%    \item We are the first to apply SSL to prostate zone segmentation [?]
\end{itemize}

\section{Method and Materials}
\label{sec:methods}
\noindent Our method uses two stages of training: (I) In a warm-up phase, a supervised model is trained with only labeled samples $T_L=(X_L, Y_L)$ until convergence, (II) semi-supervision is added to improve model performance by leveraging unlabeled images \(X_U\). At this stage, two SSL techniques are combined, namely self-learning and temporal ensembling \cite{laine2016temporal}. 
The idea behind self-learning is to get an improved model iteratively through an expanded dataset comprising the labeled dataset ($T_L$) along with the unlabeled images $X_U$ and their pseudo labels $Y_U$. To constrain the influence of wrong pseudo labels, we base on the most confident predictions using uncertainty measures \cite{Gal.2015}. We combine this idea with some concepts derived from temporal ensembling, where the pseudo-labels are updated with the ensemble predictions rather than the current epoch's prediction. Also, a consistency loss is calculated between the current and ensemble predictions, which enforces consistency between the current and the previous epochs' predictions, preventing huge gradient updates.

\begin{figure}[t]
\includegraphics[width=1\textwidth]{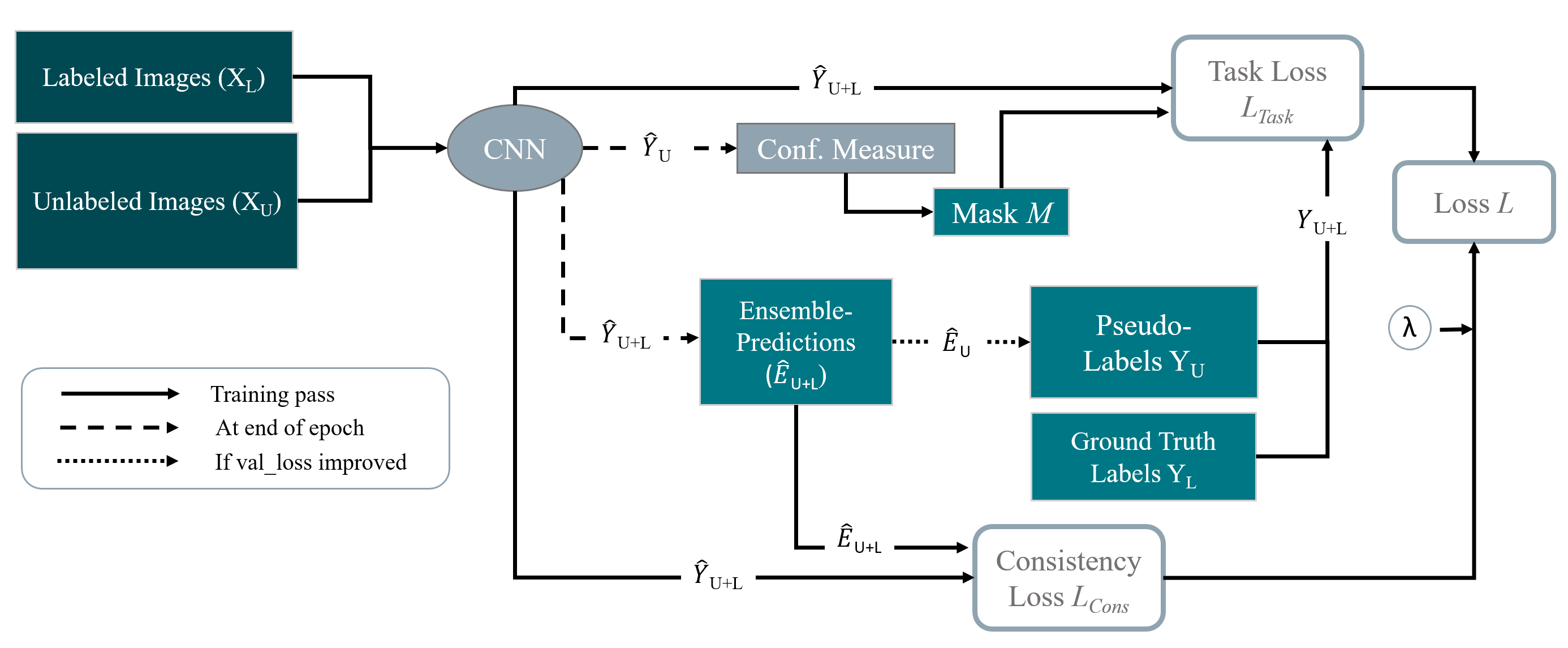}
\caption{Concept of our uncertainty aware temporal self-learning (UATS).} \label{fig:concept}
\end{figure}

\subsection{Stage I: Supervised Pre-training}
\noindent Instead of starting an SSL-based training from scratch, it is intuitive to first train a model in supervised fashion on \((X_L, Y_L)\) only, leveraging better pseudo label predictions for \(X_U\) during stage two and thus avoiding degenerated models \cite{laine2016temporal}. As our focus is not on architectural novelty but semi-supervision, we use a simple anisotropic U-Net with Dice coefficient loss as grounds for our work, which is the same architecture used as in \cite{meyer2019towards}. This combination has proven to work well for a range of biomedical segmentation tasks, including our main challenge: prostate zone segmentation. For clarity, our SSL strategy applies to all other state-of-the-art architectures.

\subsection{Stage II: Semi-supervised Learning}
\noindent An overview of our semi-supervised technique is depicted in Fig.\,\ref{fig:concept}. We start the training with the expanded dataset comprising the ground truth (manual) labels $Y_L$, and the pseudo labels $Y_U$ derived from the pre-trained model. Thus, the pseudo labels act as targets for the unlabeled samples.
%(with consecutive refinement: connected components and distance-based hole-filling, see \cite{meyer2019towards}).
The loss function \(L\) in this stage contains two components: \textit{task} and \textit{consistency} loss and is defined as weighted combination
\[L = L_{Task} + \lambda L_{Cons},\]
where $\lambda$ is the consistency loss weighting coefficient.\\

\textit{Task Loss} \(L_{Task}\) is the difference between the network output and the ground truth. We chose the continuous Dice Coefficient (cDC) \cite{Shamir.2019} to implement this loss function. It can cope with probabilistic segmentation, unlike the regular Dice Coefficient (DC) that needs at least one binary input. This exempts us from defining any thresholds on the network predictions and from losing information the network provides with its probabilistic output. Our task loss is defined as:
\begin{multline}
 L_{Task} = L_{cDC}(Y_{L \cup U}, \hat{Y}_{L \cup U},M)= \\
 -\sum_{s \in S}{\frac{2\sum_{i=1}^{N} [M_i=1] \hat{Y}_{s,i}Y_{s,i}} 
  {c_s\sum_{i=1}^{N} [M_i=1]  Y_{s,i} + \sum_{i=1}^{N} [M_i=1]\hat{Y}_{s,i}}
  }
 \end{multline}
where \(\hat{Y}\) is the model's prediction and \(Y\) is the corresponding ground truth. $N$ is the total number of voxels with $i$ being an index for a specific voxel. $S$ is the set of different structures (classes) and \(c_s\) is a specific coefficient for the cDC (see \cite{Shamir.2019} for details). Mask $M$ is one for all voxels of \(Y_L\) and for the \(n\) most confident voxels of \(Y_U\) and zero otherwise. \([ \cdot ]\) is a mask-based indicator function.

As a novelty, we propose to use the ensemble of predictions $\hat{E}$ used in the consistency loss \cite{laine2016temporal} as the pseudo labels \(Y_U\). With an ensemble of predictions, we consider multiple hypotheses rather than a probable noisy single hypothesis.
Therefore, this strategy reduces the effect of noisy labels generated for the unlabeled images. $\hat{E}$ is a weighted moving average over epochs and its update at each epoch is defined as  
\[\hat{E} \leftarrow \alpha\,\hat{E} + (1-\alpha)\,\hat{Y}\]
with $\alpha$ being defined as the momentum term controlling the contribution of historical data to the ensemble \cite{laine2016temporal}.
In contrast to the original temporal ensembling approach, we do not apply the temporal ensembling from scratch and thus do not need to correct for the startup bias. Instead, we initialize \(\hat{E}\) prior to the first  epoch with the supervised model’s prediction. 
Additionally, we propose to update \(\hat{E}\) only during those epochs where the validation loss decreases, i.e., when the model performs well on the unseen data. To be precise, \(\hat{E}\) is updated class-wise, such that the labels for the classes are updated only when the class-specific loss improved on the validation loss. 

We constrain the influence of wrong pseudo labels on the task loss by only including the $n$ most confident voxels of $\hat{E}_U$ as pseudo labels $Y_U$. We evaluate plain softmax probability output of the network and Monte Carlo (MC) dropout entropy \cite{Gal.2015} as confidence measures for the pseudo-label selection process. 
The MC dropout entropy evaluates the entropy of multiple predictions that are obtained by activating dropout at inference.
%The incorporation of dropout causes the removal of random neurons, giving rise to different networks. 
%The prediction from each of these different networks 
The predictions can be seen as MC samples from the space of all available models. The MC dropout entropy calculated using the MC samples, is used to measure the randomness of the prediction made over the same voxel across F stochastic forward passes. With $f$ denoting a forward pass, MC entropy for every voxel is defined as: 

\[ Entropy(F) = - \sum \nolimits_{s \in S} \left( \frac{1}{F} \sum \nolimits^{F}_{f=1} \hat{Y}_{s} \right) log \left( \frac{1}{F} \sum \nolimits^{F}_{f=1} \hat{Y}_{s} \right) \]

\textit{Consistency Loss} \(L_{Cons}\) is obtained by calculating the dissimilarity between the ensemble predictions \(\hat{E}_{L \cup U}\) and the current network predictions \(\hat{Y}_{L \cup U}\). Hence, $L_{Cons}$ acts as a regularizer enforcing a smoother gradient update. %The consistency loss acts as a smoothness regularization %(is kind of manifold regularization? Laine \& Aila: The difference to classical regularizers is that we induce smoothness only on the manifold of likely inputs instead of over the entire input domain)  
In the original temporal ensembling method designed for classification, the dissimilarity is measured with mean squared error. For segmentation tasks, we found it more effective to define the similarity as cDC of the two segmentations as it is less sensitive to class imbalance and can cope with probabilistic segmentations. %Furthermore, it can cope with probabilistic segmentations, unlike the regular DC that needs at least one binary input. This exempts us from defining any thresholds on the network predictions for \(\hat{E}\) and the pseudo labels and from losing information the network provides with its probabilistic output.  % TODO: if enough space is left, include example from Suhita's thesis (p. 39)
We define the consistency loss as the dissimilarity \cite{cha2007comprehensive} between $\hat{Y}$ and $\hat{E}$ as:
\[L_{Cons} = 1-L_{cDC} (\hat{Y}_{L \cup U}, \hat{E}_{L \cup U},M)\]  
 %where mask $M$ is one everywhere, because as all voxels - independent of their confidence - should be incorporated to the consistency loss.
 where mask $M$ is one everywhere, as all the voxels are considered irrespective of their confidence.

\subsection{Datasets}
\noindent Our main challenge was to solve the problem of prostate zone segmentation with the aid of unlabeled data. However, since our strategy is applicable much more widely, we decided to show its ability to generalize on other challenging tasks too. More precisely, we used the following benchmark datasets.\\

\textit{Prostate Zone Segmentation} For prostate zone segmentation we used the ProstateX data \cite{Litjens.2017,lijtens2014prostateX,Clark.2013} for which ground truth is publicly available by \cite{meyer2019towards}. The dataset comprises of multi-site prostate MRI scans with wide variety, including healthy, as well as cancer and hyperplasia affected patients. Overall, we extracted 334 T2w axial volumes, 98 of which are associated with labels for PZ, TZ, AFS and DPU zones, while the other 236 are unlabeled. To ensure comparability with \cite{meyer2019towards}, we set aside the same 20 labeled samples for testing. 
For these 20 test samples, segmentations from two clinicians (Reader\,1 and Reader\,2) are available. Another clinician (Reader\,3) segmented 10 of these cases. Segmentations from Reader\,1 (the same who segmented the training data) was set as ground truth. Segmentations from Reader\,2 and Reader\,3 were evaluated against this ground truth. The average performance of Reader 2 \& 3 forms the inter-rater level. More details on the process of ground truth creation can be found in \cite{meyer2019towards}.

The original volumes had varying resolution of $[0.3-0.6] \times [0.3-0.6] \times [3.0-5.0]$ mm. Therefore, we resampled the volumes' resolution to a common spacing of  $0.5\,\times 0.5\,\times\,3~$mm. 
% which corresponds to the highest in-plane resolution while maintaining the relation of in-plane to inter-plane resolution of the dataset.
The volumes were cropped to a unified size of $168\,\times\,168\,\times\,32$ voxels by considering the intersecting volume of the axial, sagittal and coronal T2w sequence. To be clear, we only used the cropped and resampled axial T2 volume for training. The sagittal and coronal volumes were only considered for automatically determining the region of interest. The intensities were cropped to the first and 99th percentile and subsequently normalized to an interval of [0,1]. \\

\textit{Hippocampus Segmentation} For hippocampus segmentation, we used Task04 of the Medical Decathlon Challenge \cite{Simpson.2019}. The dataset consists of two classes: the hippocampus proper (CA1-4 and dentate gyrus) and parts of the subiculum, which together are more frequently named as hippocampal formation. The T1w sagittal volumes were acquired with a Philips Achieva scanner at the Vanderbild University Medical Center (Nashville, TN, USA). The dataset consists of 390 T1w images of healthy people and patients with non-affective psychotic disorders.
For 260 out of the 390 images, labels are provided as training data. The remaining 130 samples are originally used as test data in the challenge. As these 130 images do not have any corresponding label, we randomly set aside 50 labeled samples from the original training data for our testing purposes and used the original test data  (n=130) as unlabeled (training) data for our UATS method.
The dataset's volumes are provided with uniform spacing of $1.0\,\times 1.0\,\times\,1.0~$mm for all volumes. We standardized the sizes of the volumes to $48\,\times\,64\,\times\,48$ voxels and normalized the intensities to an interval of [0,1]. \\

\textit{Skin Lesion Segmentation} For skin lesion segmentation we used the ISIC 2018 challenge data \cite{Codella.2019,Tschandl.2018}. The dataset consists of  high-resolution color photographs of the skin from all anatomic sites. It contains both benign and malignant lesions with a higher percentage of the first. Images were acquired with a variety of dermatoscope types and from different institutions. The size of the images ranges from $566 \times 679$ to $4499 \times 6748$ pixels.
The original challenge dataset provides 2594 labeled training and 1000 unlabeled testing samples. For our experiments, we used 500 randomly selected labeled samples from the original training dataset as our test data. We used the original unlabeled test data as unlabeled training data for our UATS method.  In summary, we had 2094 labeled and 1000 unlabeled images for training and 500 labelled images for testing. We employed intensity normalization and aspect-ratio normalization during preprocessing.

\subsection{Training}
%As our aim was not to achieve state-of-the-art results but to fairly compare the proposed UATS approaches with the supervised baseline. Thus, the same underlying model is used for the supervised task (Stage I) and the SSL expansion: 
\noindent We used the same training strategy for all the named tasks, but the underlying model differs slightly due to the nature of the data. For prostate zone and hippocampus segmentation, we used 3D U-Nets, while skin lesion segmentation required a 2D U-Net. In general, we used batch normalization after every convolution. The probability of the dropout in each resolution layer of the decoder path is 0.5. %A dropout of 0.5 is used in each layer of the decoder path. 
Online data augmentation included random flipping, rotation and scaling. The models were trained using ADAM \cite{kingma2014adam} optimizer based on a 75/25\% train/validation split. The hyperparameters for ADAM were set as $\beta_1$=0.9, $\beta_2$=0.999 and $\epsilon$=1e-07.

All the models were trained for a maximum of 300 epochs. We used early stopping in all the experiments, where training was stopped if there was no improvement after 30 epochs. The model with the lowest validation loss was selected and used for the evaluation on test data.
The model hyperparameters were selected empirically. In all the temporal ensembling and UATS experiments $\alpha$ was set to 0.6. The consistency coefficient was set to $\lambda = 1$ unless the consistency loss dominated the task loss. In this case, we set $\lambda = 0$ epoch-wise to ensure the task loss is the main driver. 
The percentage of confident pseudo label voxels and the Monte Carlo dropout passes were task-specific.  This originates in the different ratios of labeled and unlabeled samples among the named tasks. The task-specific hyperparameters are detailed in Tab.\,\ref{tab:hyperparam} in \ref{sec: hyperparams}.

\section{Results and Discussion}
\label{sec:evaluation} 
\subsection{Prostate Zone Segmentation}

\begin{figure*}[t!]
\centering
    \subfloat[Softmax]{\includegraphics[trim=0 600 0 600,clip,width=0.18\textwidth]{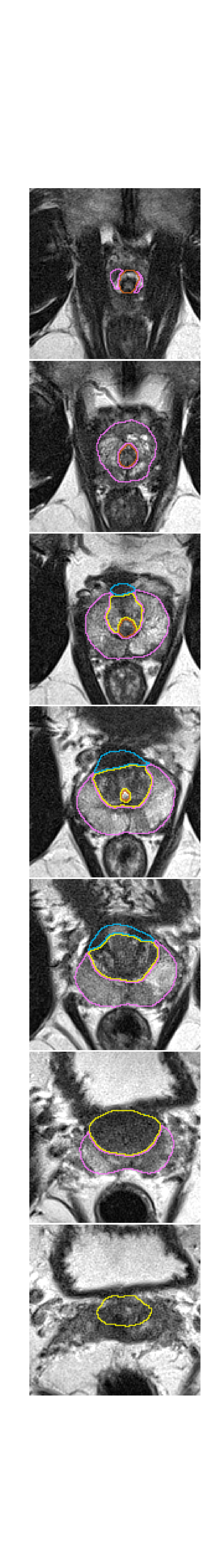}}
    \subfloat[Entropy]{\includegraphics[trim=0 600 0 600,clip,width=0.18\textwidth]{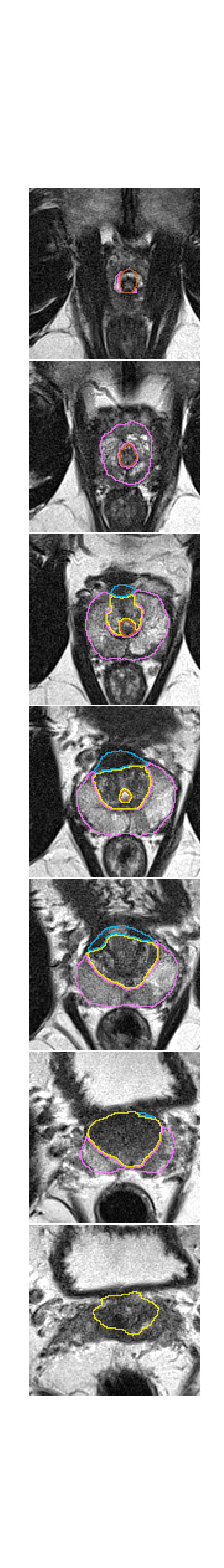}}
    \subfloat[Reader 1]{\includegraphics[trim=0 600 0 600,clip,width=0.18\textwidth]{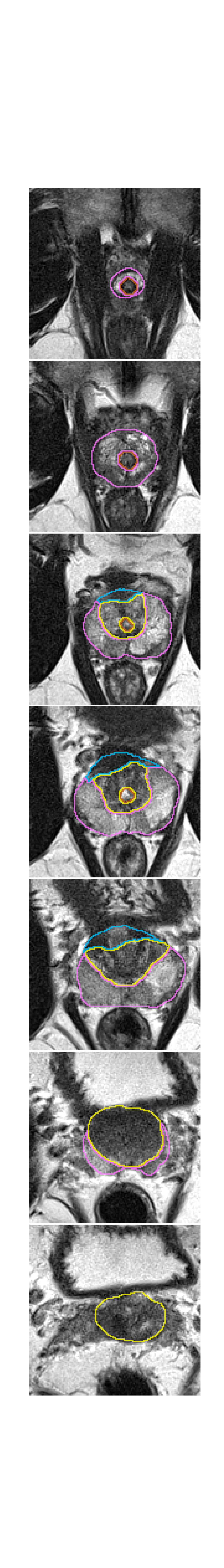}}
    \subfloat[Reader 2]{\includegraphics[trim=0 600 0 600,clip,width=0.18\textwidth]{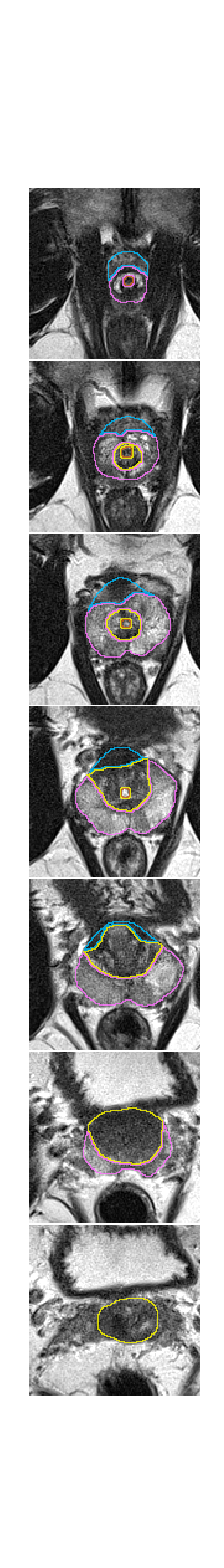}}
    \subfloat[Reader 3]{\includegraphics[trim=0 600 0 600,clip,width=0.18\textwidth]{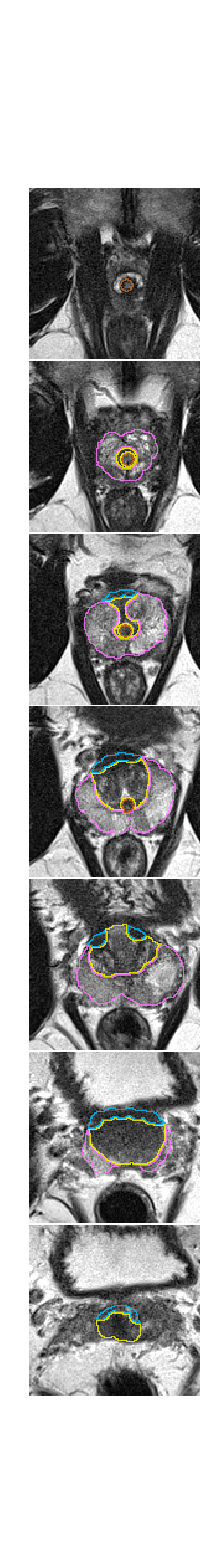}}
\caption{Example segmentation results of one test case for UATS approaches (Softmax and Entropy) and the three readers. The four structures PZ (pink), TZ (yellow), DPU (brown) and AFS (blue) are depicted.}
\label{fig:vis_res_zones}
\end{figure*}

\noindent We investigated UATS on our main prostate zone segmentation task using four-fold cross-validation and evaluated each fold on the hold-out test data with Dice coefficient (DC) and average boundary distance (ABD) as evaluation measure. Results of our experiments are summarized in Tab.\,\ref{tab:prostate}. Qualitative results for our method are visualized in Figure\,\ref{fig:vis_res_zones}.

We compared our method with Monte Carlo dropout entropy (F) and softmax outputs (G) as confidence measures to the inter-rater level (A) and the supervised baseline as in \cite{meyer2019towards} (B). We have to point out that the results for the supervised baseline (B) differ from our previous work \cite{meyer2019towards} reported in Tab.\,\ref{tab:SOTA} because in \cite{meyer2019towards} we did not use any validation data for the final model. Thus, the effective training data size was larger in\,\cite{meyer2019towards}. However, the method and hyperparameters are the same for both training procedures.
Furthermore, we evaluated the performance of temporal ensembling as in Laine and Aila~\cite{laine2016temporal} with cDC as consistency loss (C) and pseudo label self-learning as in Bai \textit{et al.}~\cite{Bai.2017} (D). Bai \textit{et al.} only update the pseudo label after a specific interval of 50 epochs regardless of the quality of the model concerning the validation data. Therefore, pseudo labels quality may decrease and the method may suffer from these low-quality labels. Thus, we apply a variant that updates the pseudo labels always and only when the validation loss improves (E).

Moreover, we investigated the impact of the main components of our proposed method in an ablation study. To be precise, we evaluated the contribution of the consistency loss (H), the uncertainty guidance (I), and the improved pseudo label that is extracted from a temporal ensemble of predictions (J). For these experiments, we only considered the UATS softmax variant. We refer to the appendix for more detailed results for the prostate zone segmentation experiments (see \ref{app:zones_results} and \ref{app:ablation}).

In another experiment, we compared how robust against noise the supervised baseline (B) and the UATS approaches (F and G) are. For this, we applied varying noise levels on the test data and evaluated the performance of the methods.

\subsubsection{Comparison to the State-of-the-Art}

\begin{figure}[t]
\centering\includegraphics[width=\textwidth]{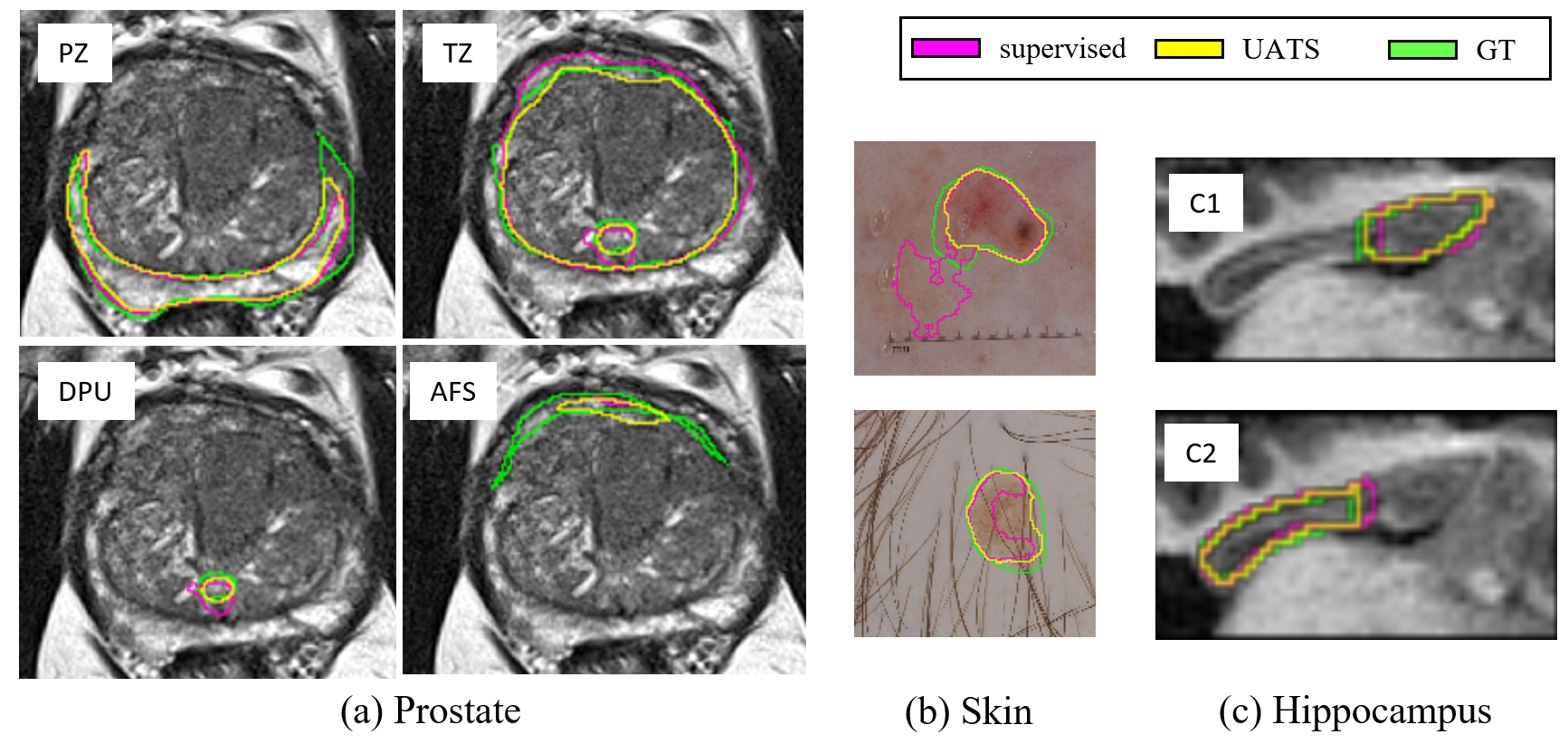}
\caption{Qualitative Results visualizing the improvement from supervised to UATS for prostate, skin and hippocampus segmentation.} \label{fig:vis_results}
\end{figure}

\noindent As summarized in Tab.\,\ref{tab:prostate}, UATS (F and G) outperformed all the other methods (B-E) with most performance gain for the minority classes DPU and AFS. These are also the challenging classes, as can be seen from the low inter-rater quality. The challenge arises from these structures' heterogeneous appearance and shape, making them profit from the additional samples. The other two classes, PZ and TZ, gain only little from semi-supervision, irrespective of the method considered. 
Fig.\,\ref{fig:vis_results}a) shows the improvement UATS obtained over the supervised baseline for one case. In this example, every zone benefits from the semi-supervised addition, leading to closer predictions to the ground truth.

\begin{table}[t]
\centering
\caption{Dice coefficients (DC) and average boundary distances (ABD) of different prostate zone segmentation strategies. Best results are marked bold. Asterisks mark significantly better results when compared to the supervised baseline (according to Wilcoxon signed-rank test).}
\label{tab:prostate}
    \resizebox{1\linewidth}{!}{%
 \begin{tabular}{l|ll|ll|ll|ll}
     \textbf{Algorithm}& \multicolumn{2}{l}{\textbf{PZ}} & \multicolumn{2}{l}{\textbf{TZ}} & \multicolumn{2}{l}{\textbf{DPU}} & \multicolumn{2}{l}{\textbf{AFS}} \\
           & DC (\%)        & ABD (mm)        & DC (\%)         & ABD (mm)      & DC (\%)         & ABD  (mm)      & DC (\%)          & ABD  (mm)      \\
  \hline
  \hline

(A) Inter-rater-level (from \cite{meyer2019towards}) & $79.9\pm4.2$ & $0.857 \pm 0.451$&  $85.3\pm5.8$ & $0.969 \pm 0.320$ & $62.4\pm7.8$ & $1.297 \pm 0.460$ & $48.9\pm12.6$ & $2.164 \pm 1.152$ \\
(B) Supervised & $77.4 \pm 5.7$ & $1.079 \pm 0.671$ & $86.7 \pm 7.2$ & $0.911 \pm 0.363$ & $70.6 \pm 14.7$ & $1.079 \pm 1.319$ & $46.1 \pm 13.6$ & $3.547 \pm 3.178$\\
(C) Temp. Ens. & $77.4 \pm 6.1$ & $1.183 \pm 0.893$ & $87.0 \pm 6.5$ & $0.931 \pm 0.416$ & $71.6 \pm 12.0$ & $0.953 \pm 0.564$ & $44.2 \pm 11.9$ & $3.275 \pm 2.773$\\
(D) Self-Learning & $77.1 \pm 5.9$ & $1.127 \pm 0.817$ & $84.4 \pm 6.4$ & $1.136 \pm 0.459$ & $68.8 \pm 13.2$ & $1.068 \pm 0.695^{***}$ & $43.4 \pm 11.2$ & $3.407 \pm 2.300$\\
(E) Pseudo-Update & $77.0 \pm 5.6$ & $1.290 \pm 0.983$ & $85.4 \pm 6.0$ & $1.050 \pm 0.366$ & $71.2 \pm 11.0$ & $0.970 \pm 0.634$ & $41.2 \pm 12.9$ & $4.421 \pm 3.384$\\
\hline
(F) UATS Entropy & $\textbf{78.9} \pm \textbf{5.0}^{***}$ & $\textbf{0.973} \pm \textbf{0.59}6^{***}$ & $\textbf{87.3} \pm \textbf{6.5}^{**}$ & $\textbf{0.889} \pm \textbf{0.359}$ & $73.6 \pm 9.9^{***}$ & $0.872 \pm 0.463^{**}$ & $50.1 \pm 10.3^{***}$ & $\textbf{2.948} \pm \textbf{2.046}^{*}$\\
(G) UATS Softmax & $78.6 \pm 5.4^{***}$ & $1.024 \pm 0.660^{**}$ & $86.9 \pm 6.6$ & $0.936 \pm 0.387$ & $\textbf{75.3} \pm \textbf{9.4}^{***}$ & $\textbf{0.811} \pm \textbf{0.420}^{***}$ & $\textbf{50.6} \pm \textbf{11.7}^{***}$ & $2.991 \pm 2.244^{*}$\\
\hline
(H) UATS w/o $L_{Cons}$ & $78.4 \pm 5.6^{***}$ & $1.026 \pm 0.673^{*}$ & $86.8 \pm 6.3$ & $0.926 \pm 0.334$ & $73.2 \pm 9.4^{*}$ & $0.898 \pm 0.451$ & $47.0 \pm 12.3$ & $3.563 \pm 2.505$\\
(I) UATS w/o confidence & $77.3 \pm 5.6$ & $1.088 \pm 0.692$ & $86.6 \pm 6.9$ & $0.945 \pm 0.402$ & $70.9 \pm 12.2$ & $0.978 \pm 0.585$ & $47.8 \pm 11.9$ & $3.455 \pm 2.604$\\
(J) UATS w/o ensemble for PL & $78.4 \pm 5.6^{***}$ & $1.086 \pm 0.750$ & $86.9 \pm 6.6$ & $0.933 \pm 0.383$ & $75.0 \pm 8.8^{***}$ & $0.818 \pm 0.369^{***}$ & $49.1 \pm 11.5^{***}$ & $3.231 \pm 2.467$\\
\bottomrule
\multicolumn{4}{l}{${}^*p<0.05$, ${}^{**}p<0.01$, ${}^{***}p<0.001$.}
\end{tabular}}
\end{table}

% This is not surprising since the supervised baseline is already close to the inter-rater level, indicating that the labeled samples already cover their variability well enough. It is also noticeable that UATS is the only method being on par with the inter-rater level for all zones.

With respect to the inter-rater performance (A), our UATS' performance is in the range of the inter-rater level for all structures. The UATS segmentations yield even higher DC for all structures except PZ. We carried out further analysis of the model results for PZ and examined in which region most of the automatic and expert segmentation disagreement occur. We found that most deviations from the inter-rater level can be encountered in the upper (cranial) third of the PZ, which is located in the prostate’s base in proximity to the seminal vesicles (SV). This region suffers from partial volume effect (PVE) due to the high slice thickness and the intensity similarity of PZ and SV. While the automatic segmentation only processes the (anisotropic) transversal T2w scan, the human readers could additionally verify their segmentations in the sagittal T2w scans. In the sagittal scan, the SVs can be better distinguished from PZ tissue. We assume that this is one reason why automatic performance is lower than the inter-rater-level for PZ. Another reason could be that, for PZ, the amount of labeled data needs to be increased to better cover the structure's variety.

The above statements are valid for both UATS confidence measures: softmax probability and Monte Carlo dropout entropy. We would recommend the softmax probability for this task. This is because both performances are about equal and the softmax probability is cheaper to compute.

It is interesting to see that the relatively simple pseudo labeling approaches (D and E) generally lead to a performance decline compared to the supervised baseline (B) for most structures. This demonstrates that for the prostate dataset, naive self-learning setups suffer from the confirmation bias reported in the literature \cite{arazo.2020}.
However, in combination with the uncertainty guidance and  consistency loss, we see that our UATS method improves significantly over the supervised baseline.

\subsubsection{Ablation Study of UATS Components}
\noindent 
We conducted an ablation study to investigate the individual effect of different components on the overall performance. For this, we employ the UATS softmax variant and make the following changes to its implementation: 
First, we omitted the consistency loss $L_{cons}$ and only optimized the network parameters during training with $L_{Task}$ (Experiment (H)).
Second, we set the parameter for the percentage of confident voxels to $n=100\%$. This way, \textit{all} pseudo label voxels are considered as confident and not only the top $n$ voxels (with highest probability) per class. Thus, no confidence measure is included in the UATS method (Experiment (I)).
And third, instead of an ensemble of predictions \(\hat{E}\), we use the current prediction of the network \(\hat{Y}\) as pseudo labels \(Y_U\) (Experiment (J)).

Experiment (H) demonstrates the decreased performance when the $L_{Cons}$ is omitted and the network is only trained with $L_{Task}$. This implies that adding the consistency loss to the uncertainty-guided self-learning in the UATS setting, improves the method’s accuracy.
The improvement accounts, in particular, for the smaller structures. Laine and Aila \cite{laine2016temporal} demonstrated that temporal ensembling could better cope with noisy labels than a simply supervised baseline. We can assume that this has a more massive effect on smaller labels because the ratio of falsely labeled voxels is larger when the structure's total size is relatively small.

By only considering the $n$ most confident voxels of each class, we mitigate the likelihood of falsely labeled voxels in our pseudo labels. If we consider all pseudo voxels for the labels (I), we see an evident performance decline for almost all structures in comparison to our proposed UATS Softmax (G). %Even compared to the supervised baseline, the performance is worse for all structures except for AFS. 
This demonstrates that for a beneficial effect of self-learning, the choice of labels  incorporated into the task loss is essential. This finding is in line with many studies that implement self-learning and some form of uncertainty guidance (e.g.\,\cite{Nie.2018}).

In contrast to the consistency loss and confidence guidance, the ensemble of predictions for pseudo labels (J) does not  contribute to the improvement as the confidence and the consistency loss. We hypothesize that selecting the most confident voxels already reduces the false voxel label's contribution, and the consistency loss compensates further for the false voxels.\\

\subsubsection{Robustness against Noise}

\begin{figure}[b]
\centering
\includegraphics[width=1.0\textwidth]{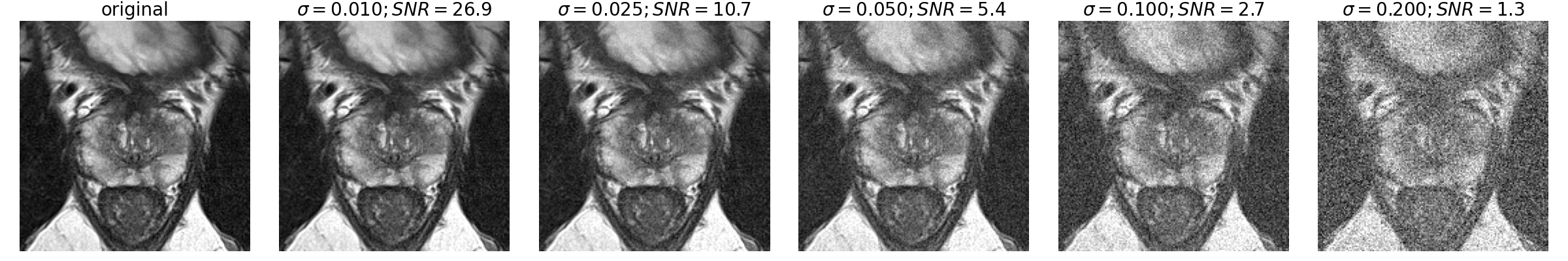}
\caption{Application of additive Gaussian noise with varying $\sigma$ to one example test case.} 
\label{fig:noise_vis}
\end{figure}

\begin{figure}[t]
\centering
\subfloat[]{\includegraphics[width = 0.25\textwidth]{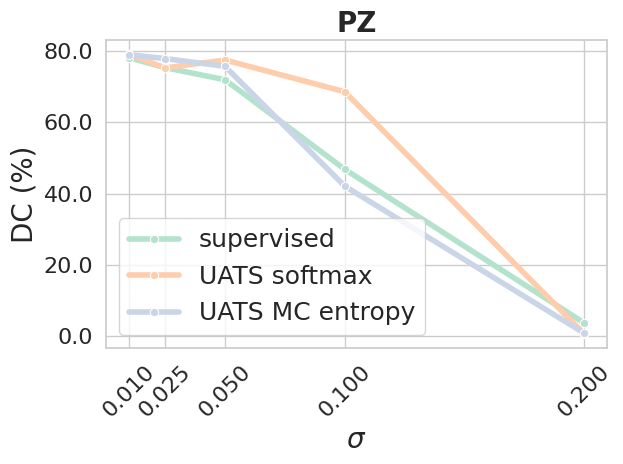}} 
\subfloat[]{\includegraphics[width = 0.25\textwidth]{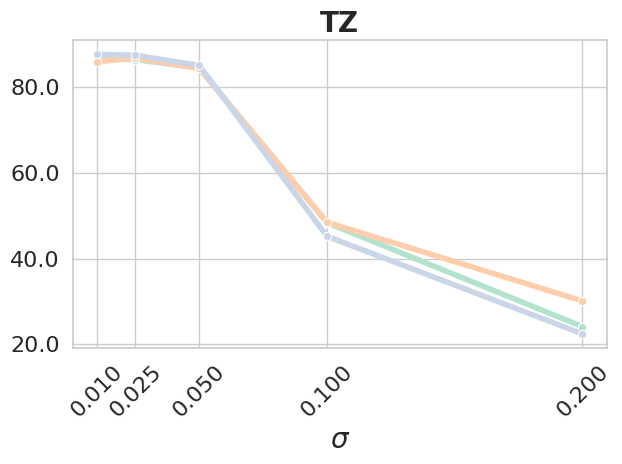}} 
\subfloat[]{\includegraphics[width = 0.25\textwidth]{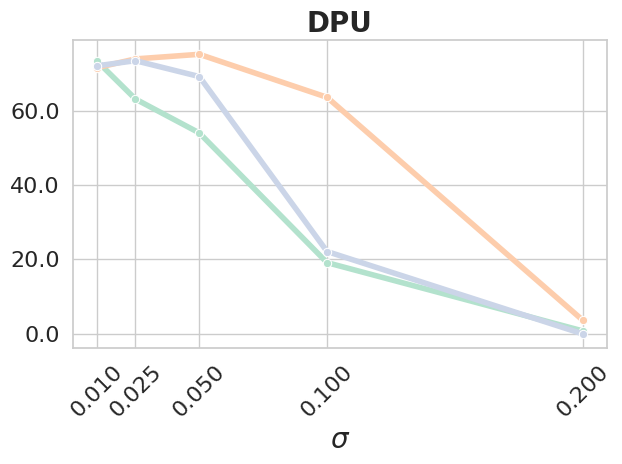}} 
\subfloat[]{\includegraphics[width = 0.25\textwidth]{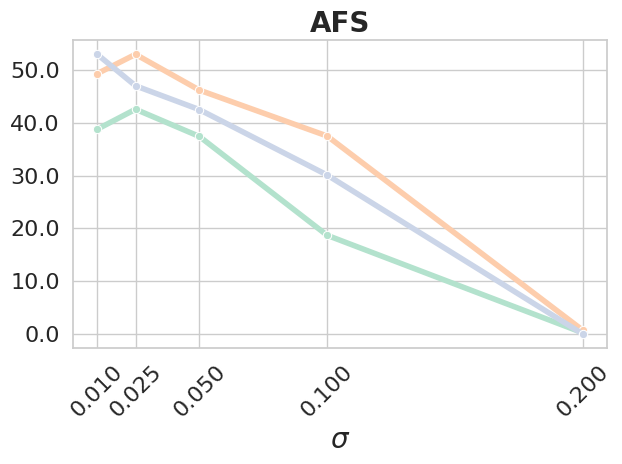}} 
\\
\subfloat[]{\includegraphics[width = 0.25\textwidth]{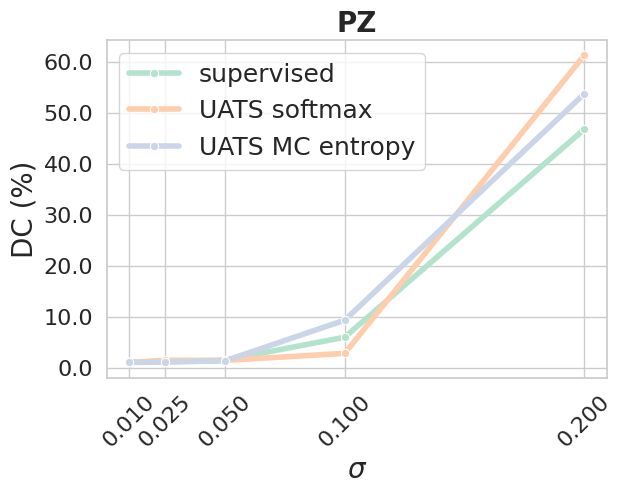}} 
\subfloat[]{\includegraphics[width = 0.25\textwidth]{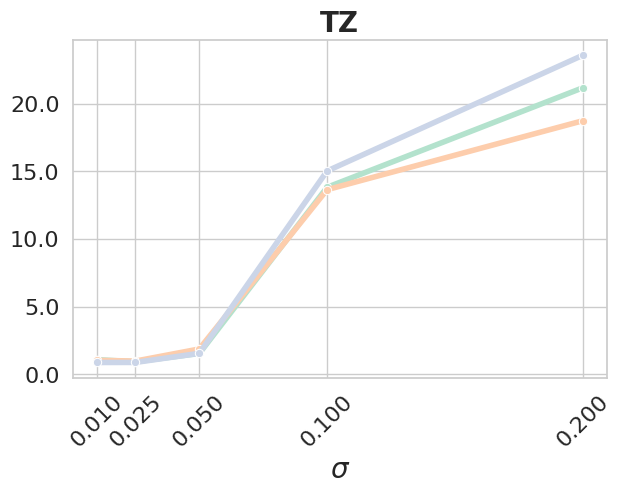}} 
\subfloat[]{\includegraphics[width = 0.25\textwidth]{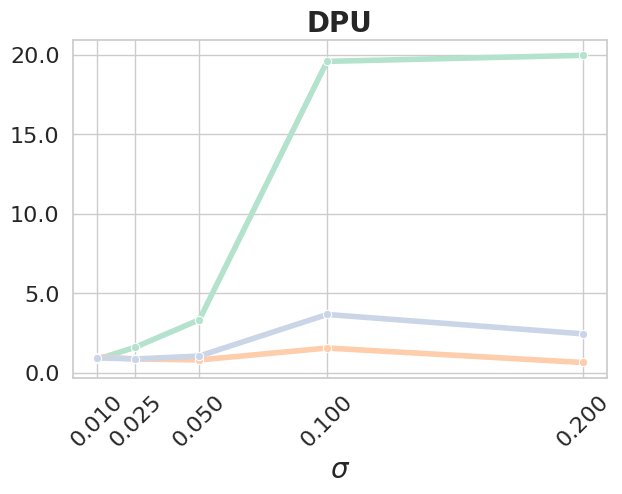}} 
\subfloat[]{\includegraphics[width = 0.25\textwidth]{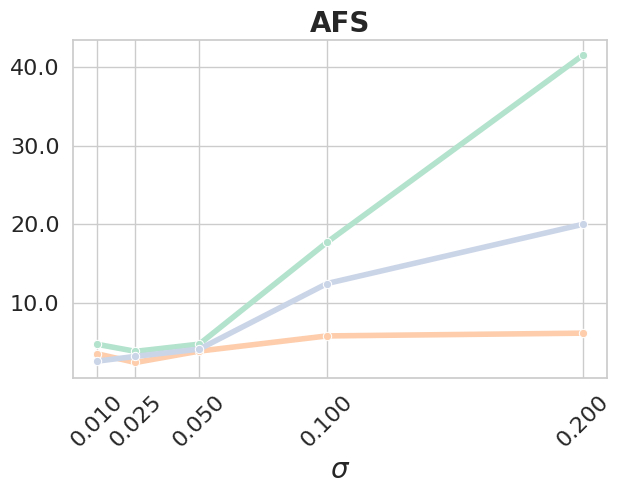}} 
\caption{Performance of the supervised baseline with UATS softmax and UATS entropy on the test dataset with added Gaussian noise with varying $\sigma$. Performance measures are DC a)-d) and ABD e)-h).} 
\label{fig:plots_noise}
\end{figure}

In this experimental setting, we investigated the robustness of the proposed algorithm against varying levels of noise. We applied Gaussian additive noise with $\mu=0$ and varying $\sigma$  to our test images. Because MRI images have different intensity ranges, we applied the noise to the normalized test image. We normalized the noisy images again to obtain the same intensity range from [0,1] as for the training images. We applied noise with $\sigma=\{0.01, 0.025, 0.05, 0.1, 0.2\}$ which corresponds to an average signal-to-noise-ratio ($SNR=\frac{\mu_{img}}{\sigma_{noise}}$) of $SNR=\{26.8, 10.7, 5.4, 2.7, 1.3\}$.
An example case with increasing noise is displayed in Figure \ref{fig:noise_vis}.

We evaluated the performance of the supervised baseline, the UATS softmax and the UATS entropy approach. The results are plotted for DC and ABD in Figure \ref{fig:plots_noise}. The general performance drops for all approaches with noise strength above $\sigma=0.05$. For TZ and PZ, all three approaches perform similar. For the smaller structures AFS and DPU, however, the UATS softmax suffers much less from noise, indicating its increased robustness against noise compared to the other two approaches. 

We did not apply any intensity augmentation in our training. Consequently, we assume that all methods would generally improve their performance with increasing noise, if they had seen it during training.

Besides noise, there are various other factors that impact the quality of images. For example, images can be blurry due to motion of the patient during acquisition or they can be corrupted by bias field.  More intensity-based data augmentation and more task-specific preprocessing could but does not guarantee to increase the robustness of all algorithms. The majority of state-of-the art approaches uses established preprocessing consisting of intensity normalization and spatial normalization (image resolution and size). We followed this style for our work, as our focus was to investigate how additional unlabeled data can improve over a supervised baseline. 

The image resolution is another important parameter for the performance of automatic but also manual segmentations.  If the resolution is very low, the images suffer from higher PVE. This is for example the case for the apex and base region of the prostate in highly anisotropic transversal T2w images. Here, the structure boundaries are more difficult to distinguish than in the mid-gland region.

\subsection{Generalizability across Tasks}
\label{varying_labeled}
\begin{figure}[t]
\centering
\includegraphics[width=\textwidth]{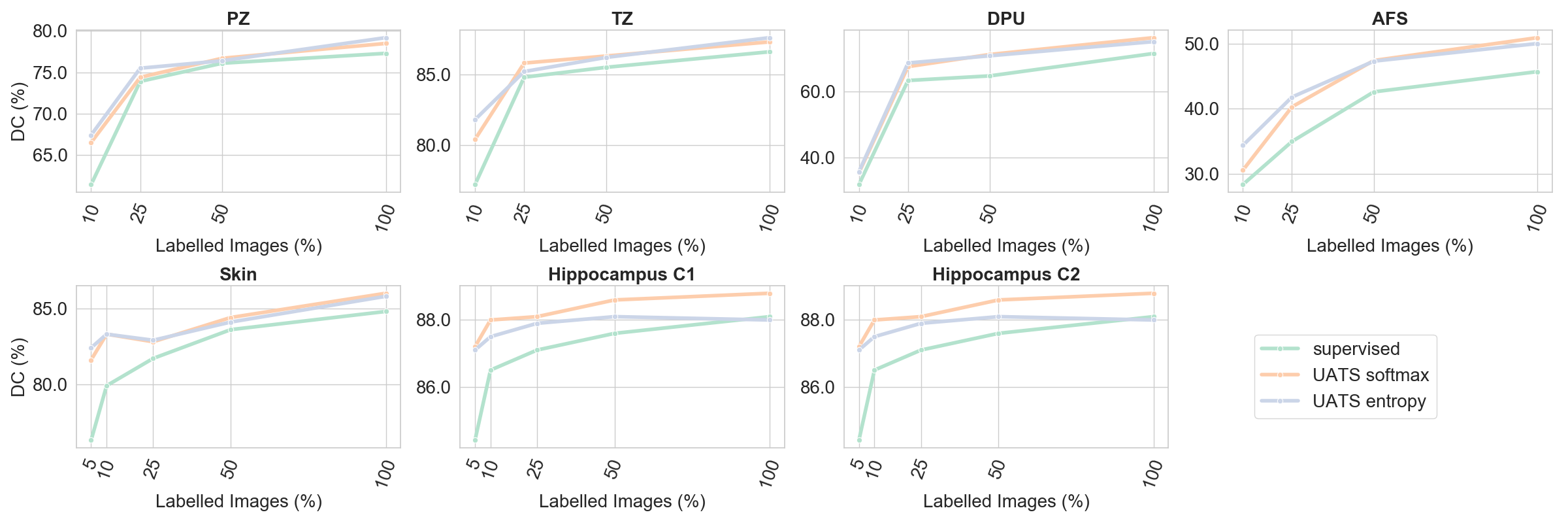}
\caption{Supervised and UATS methods' performance (DC) depending on the amount of labeled samples for the tasks of prostate zone, skin lesion and hippocampus segmentation.}
%[In x-axis lieber ratio of labeled/unlabeled? same diagrams for all prostate zones?]}
\label{fig:results_diagram}
\end{figure}

\noindent To show that UATS is generally applicable, we additionally benchmarked it on a hippocampus and skin lesion segmentation dataset and for varying amounts of labeled data. We set up the experimental scheme for all datasets as follows. We randomly sampled 5, 10, 25, 50 and 100 \% of the available labeled samples and applied the plain supervised baseline (Stage I) and UATS (Stage II), whereby we used all the available unlabeled samples. The random sampling was repeated three times, averaging result qualities to get reasonable grounds for comparison. Fig.\,\ref{fig:vis_results} (b) and (c) gives a visual comparison of the methods for skin and hippocampus data. Quantitative results of our experiment are presented in Fig.\,\ref{fig:results_diagram}.

Reviewing the results, one can conclude that UATS outperformed the supervised baseline irrespective of the dataset and the number of labeled samples. The general tendency is that the smaller the amount of labeled samples, the larger the gain from unlabeled samples. This is a common observation in SSL methods, also mentioned in \cite{Bai.2017}.  %The performance of UATS over the supervised baseline is highest for smaller amounts of labeled training data. 
The rationale is quite intuitive, at some point, the variability appearance and shape are well-covered by the labeled samples, yielding diminishing returns from unlabeled samples.
The only exception from this intuitive finding is the DPU segmentation with 10\,\%  of labeled samples, where the gain from UATS is small. Presumably, this is caused by the fact that 10 \% equals 6 labeled samples, which might generally be insufficient for a reasonable DPU segmentation.

The above findings are true for both UATS confidence measures. However, the results also show that both confidence measures are not equivalent: for the prostate zone and skin lesion segmentation, they perform about equal, but for hippocampus segmentation, softmax probability outperforms Monte Carlo dropout entropy over a wide range of labeled samples.

\subsection{Comparison to Others}
\noindent Above, we compared UATS with the supervised baseline from \cite{meyer2019towards} and showed that it significantly increased the performance. As summarized in Tab. \ref{tab:SOTA}, various approaches have been proposed for the segmentation of only PZ and TZ. When comparing our UATS method with other approaches, UATS performs in the mid-range. But a direct comparison to other methods can not be drawn for several reasons. First,  we target a more detailed anatomy inside the prostate such that our problem definition is more difficult than the two class segmentation of PZ and TZ. And second, the underlying datasets are not the same. They differ in sample size, acquisition protocols and image quality. 
%And third, the evaluation procedure is not the same. Some works evaluate the algorithm outcome on the validation data and others on hold-out test data.
%An alternate approach based on spatial pyramid networks for PZ and TZ segmentation was presented in \cite{Liu.2019}. Our method also outperforms their segmentation results on ProstateX data (TZ: 79\% DC and PZ: 74\% DC).

For the ISIC2018 skin lesion segmentation challenge, hundreds of methods are listed in the live leader-board, and the currently best one achieves a DC of 92.2\% while our UATS method obtained a DC of 86.0\%. A lot of algorithms are not accompanied by a method description, and thus we can only assume why there is a performance gap between our and the top participants.

One reason might be the varying amounts of training data. As we set 500 labeled samples for test purposes aside, we have 500 labeled samples less for training. Furthermore, the models for the challenge are maximally fine-tuned to this specific task. Additionally, an ensemble of models, more sophisticated preprocessing and augmentation strategies, and more complex network architectures may have been used. Our method does support such techniques, but it is beyond the scope of our work to implement them.

The Medical Decathlon challenge focuses on developing methods that have high generalization capacity on different segmentation tasks. We followed a similar strategy, and thus our results are more comparable to this challenge than the ISIC2018. In the ongoing Medical Decathlon challenge, the hippocampus's best result achieved a DC of 90.0\% for class 1 and DC of 89.0\% for class 2. With UATS, we achieved a DC of 88.8\% and 87.6\%, respectively. This places our method above the eighth-best result (88\% and 87\%), whereas we had less labeled data for training (with the same reason as  ISIC2018).

\section{Conclusion}
\label{sec:conclusion}
\noindent We proposed a semi-supervised method for prostate zone segmentation from T2w MRI. Our method combines uncertainty-aware self-learning and temporal ensembling into a novel framework to improve supervised deep learning models by commonly available unlabeled data. Regarding prostate zone segmentation, our method yields results from the quality on an inter-rater level. The improved segmentation quality of the prostate zones may enable more precise and consistent lesion location assignment, improved cancer therapy planning and could increase the accuracy of automatic lesion detection and staging methods. %, indicating that the task can be considered solved when exploiting unlabeled samples.

We showed that our method increases robustness against noise compared to the supervised baseline. Moreover, we demonstrated that UATS generalizes to other tasks by evaluating additional biomedical challenge datasets with varying amounts of labeled samples. Our experiments demonstrate that our method improves upon the supervised baselines for different ratios of labeled and unlabeled samples and different tasks. We used standard U-Nets as supervised grounds for comparison because these have demonstrated their potential for a wide range of tasks. However, our semi-supervised strategy applies to network architectures beyond U-Nets, too.
Therefore, our approach can have an impact on many different (biomedical) segmentation tasks by reducing the amount of necessary labeled images.

We found that, when gains from semi-supervision are larger, the higher the variability in appearance and shape, and the smaller the amount of labeled samples. We also found that when enough labeled samples with sufficient quality become available,  gains from semi-supervision will diminish at some point. %We observed that the more uncertain the supervised model is, the more gain from the UATS was detected.

In the future, we want to examine the robustness of the automatic zone segmentation on a larger test dataset with detailed information about the prostate being affected by hyperplasia, prostatitis, cysts or different stages of cancer. It should also be analyzed how much systems for automatic PCa detection can benefit from the the automatic zone segmentation.
Furthermore, we want to investigate whether a relation can be quantified automatically for a given task to answer the following two questions. First, can we know beforehand that semi-supervision will help? Second, can we estimate beforehand how much semi-supervision will help? For example, approaches that measure the model uncertainty \cite{Gal.2015, mehrtash2020} and estimate the segmentation quality \cite{Robinson.2018} could be investigated to address these research questions. Moreover, we would be interested to examine whether the approach could gain improvement on other biomedical imaging tasks, as for example classification.



%%       \citet{<label>} ==> Jones et al. [21]
%%       \citep{<label>} ==> [21]
%%

\section*{Conflict of Interest}
We declare no conflict of interest.

\section*{Acknowledgement}
This work has been funded by the EU and the federal state of Saxony-Anhalt, Germany under grant number ZS/2016/08/80388. This work was in part conducted within the context of the International Graduate School MEMoRIAL at Otto von Guericke University (OVGU) Magdeburg, Germany, kindly supported by the European Structural and Investment Funds (ESF) under the programme "Sachsen-Anhalt WISSENSCHAFT Internationalisierung“ (project number ZS/2016/08/80646). The funders had no role in the study design, data collection and analysis, decision to publish, or preparation of this manuscript.
The Titan Xp used for this research was donated by the NVIDIA Corporation. Data used in this research were obtained from The Cancer Imaging Archive (TCIA) sponsored by the SPIE, NCI/NIH, AAPM, and Radboud University.
\bibliographystyle{elsarticle-harv}





% \end{thebibliography}

\pagebreak
\appendix

\section{Dataset specific hyper-parameters} 
\label{sec: hyperparams}
%The dataset specific hyperparameters have been detailed in the table \ref{tab:hyperparam}.
\begin{table}[hp]
\centering
%\caption{Dataset specific hyperparameters}
    \resizebox{\linewidth}{!}{%
    \begin{tabular}{l|p{2cm}|p{2cm}|p{3cm}|l}
    \hline
    \hline
    \textbf{Dataset} &\textbf{Learning \newline Rate}&  \textbf{Batch Size} & \textbf{Number of MC Forward passes (F)} & \textbf{Confident pseudo label voxels per class}  \\
    \hline
    Prostate & 5e-5 & 2  &10 & PZ: 50\%, TZ: 50\%, DPU: 10\%, AFS: 10\%, Background: 50\% \\
     \hline
    Skin & 1e-5 & 8  &10 & Lesion: 50\%, Background: 50\% \\
    \hline
    Hippocampus & 4e-5 & 4 &10 & C1: 50\%, C2: 50\%, Background: 50\% \\
    \bottomrule

\end{tabular}}
\caption{Task-specific hyperparameters settings in our experiments.}
\label{tab:hyperparam}
\end{table}

\pagebreak

\section{Results for Prostate Zone Segmentation}
\label{app:zones_results}
The figures in this section visualize the boxplots for the prostate zone segmentation experiment for UATS and the other state-of-the-art techniques for DC and the average boundary distance (ABD).

\begin{figure}[hp]
\centering
\subfloat[]{\includegraphics[width = 0.5\textwidth]{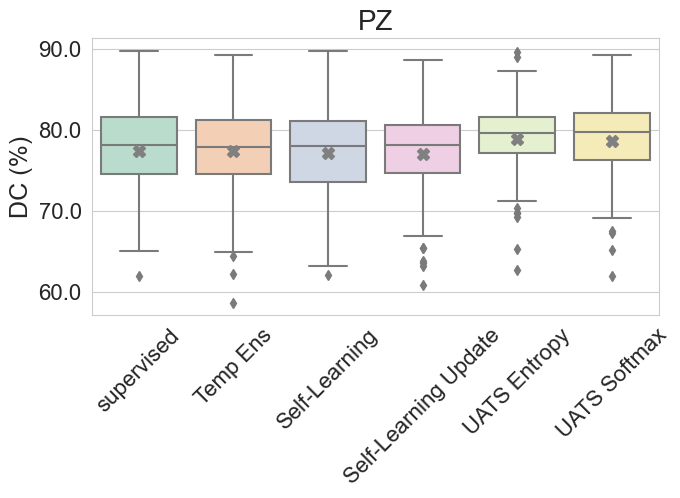}} 
\subfloat[]{\includegraphics[width = 0.5\textwidth]{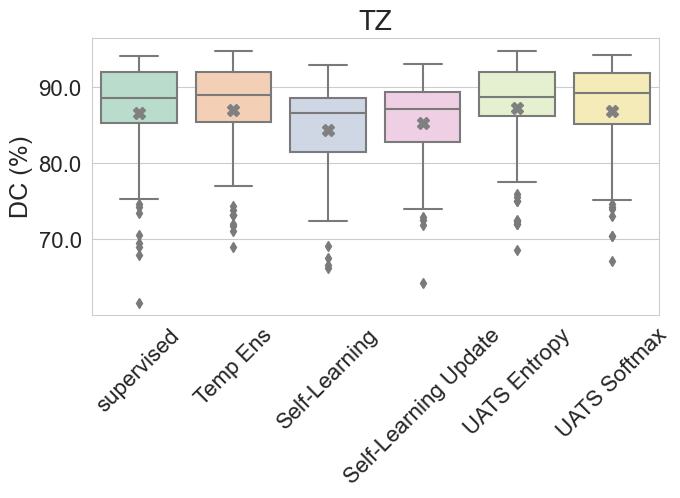}} \\
\subfloat[]{\includegraphics[width = 0.5\textwidth]{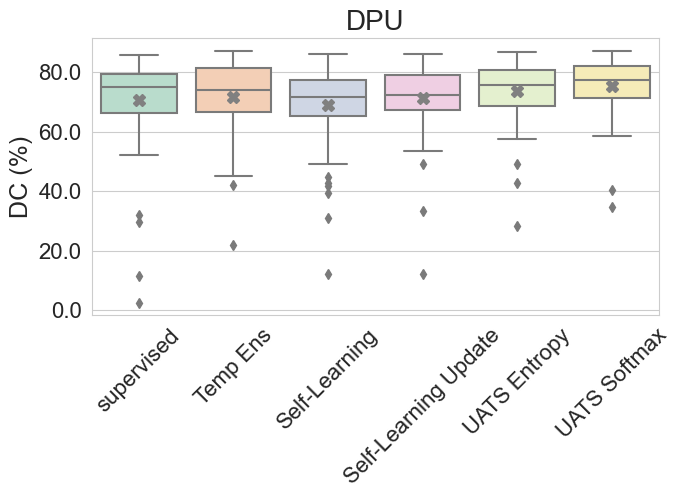}} 
\subfloat[]{\includegraphics[width = 0.5\textwidth]{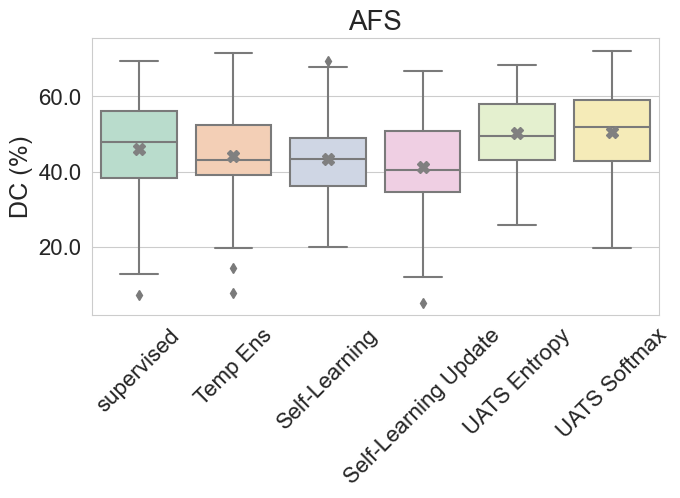}} 

\caption{Boxplots for the segmentation results of the four zones of the prostate. Results are given as the DC (\%) of the ground truth and automatic segmentation. The state-of-the-art methods are shown in comparison to our proposed UATS method.} 
\label{fig:boxplots_dice}
\end{figure}

\begin{figure}[hp]
\centering
\subfloat[]{\includegraphics[width = 0.5\textwidth]{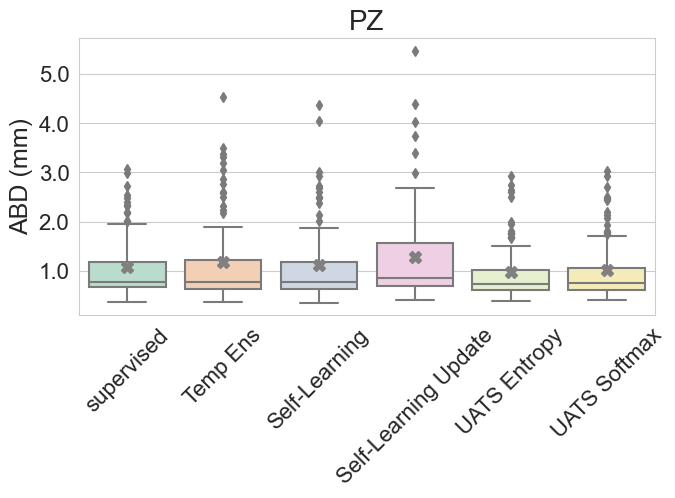}} 
\subfloat[]{\includegraphics[width = 0.5\textwidth]{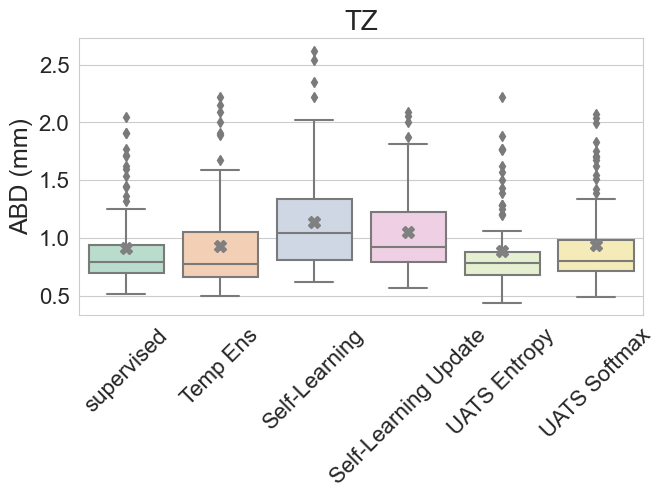}} \\
\subfloat[]{\includegraphics[width = 0.5\textwidth]{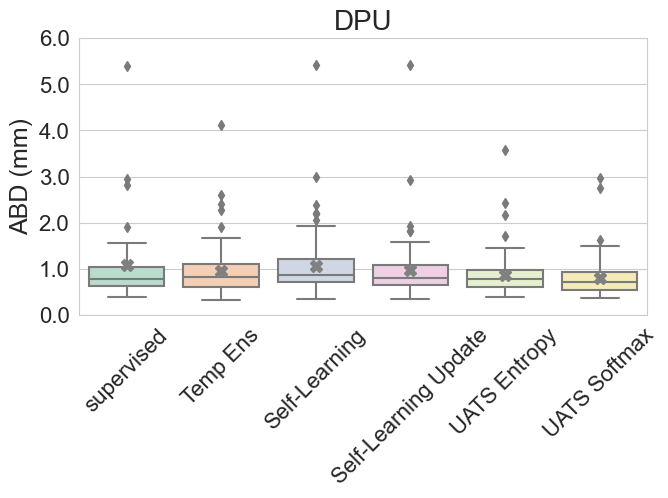}} 
\subfloat[]{\includegraphics[width = 0.5\textwidth]{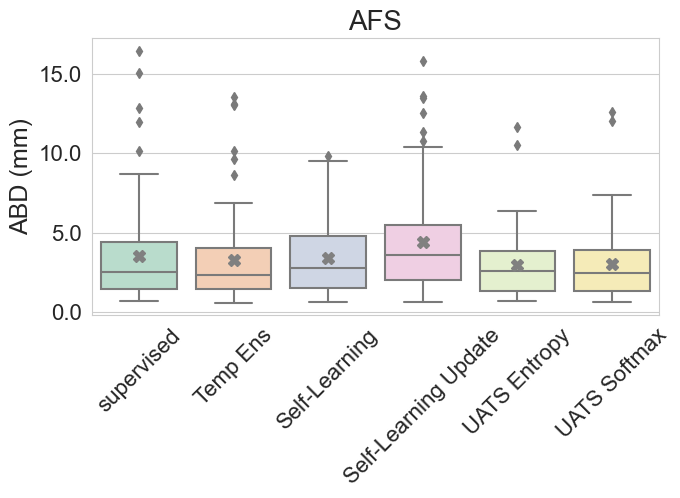}} 

\caption{Boxplots for the segmentation results of the four zones of the prostate. Results are given as the average boundary distance (ABD) in $mm$ between the ground truth and automatic segmentation. The state-of-the-art methods are shown in comparison to our proposed UATS method.} 
\label{fig:boxplots_abd}
\end{figure}

% boxplots for ablation study
\pagebreak

\section{Results for Ablation Study}
\label{app:ablation}
The figures in this section visualize the boxplots for ablation study for the UATS method with DC and ABD as evaluation measures.

\begin{figure}[hp]
\centering
\subfloat[]{\includegraphics[width = 0.5\textwidth]{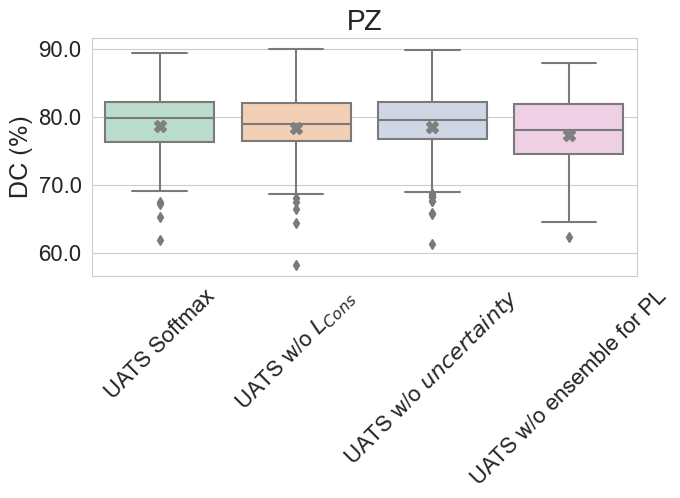}} 
\subfloat[]{\includegraphics[width = 0.5\textwidth]{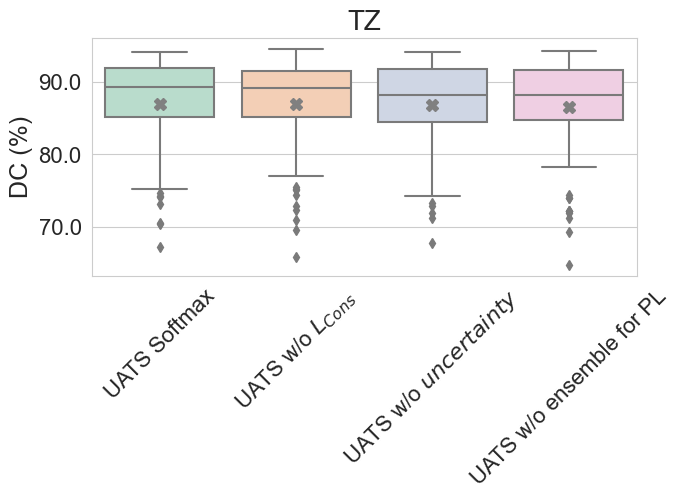}} \\
\subfloat[]{\includegraphics[width = 0.5\textwidth]{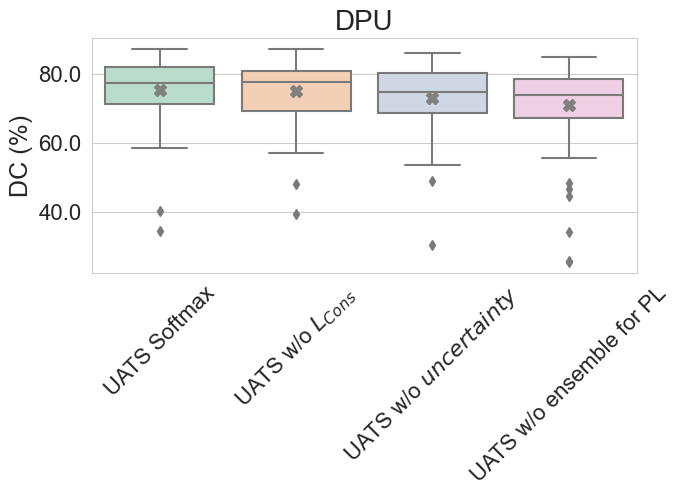}} 
\subfloat[]{\includegraphics[width = 0.5\textwidth]{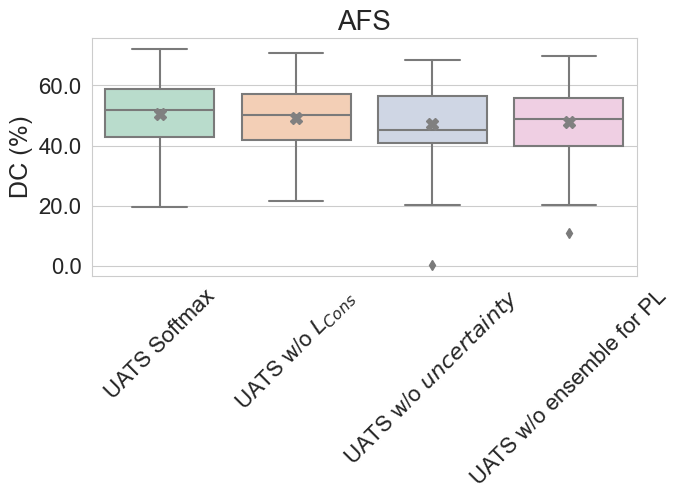}} 

\caption{Boxplots for the segmentation results of the ablation study for UATS. Results are given as the DC of the ground truth and automatic segmentation. The state-of-the-art methods are shown in comparison to our proposed UATS method.} 
\label{fig:boxplots_ablation_dice}
\end{figure}

\begin{figure}[hp]
\centering
\subfloat[]{\includegraphics[width = 0.5\textwidth]{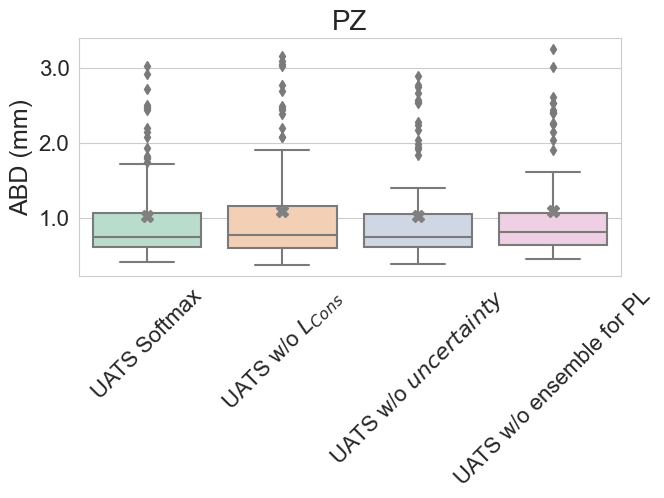}} 
\subfloat[]{\includegraphics[width = 0.5\textwidth]{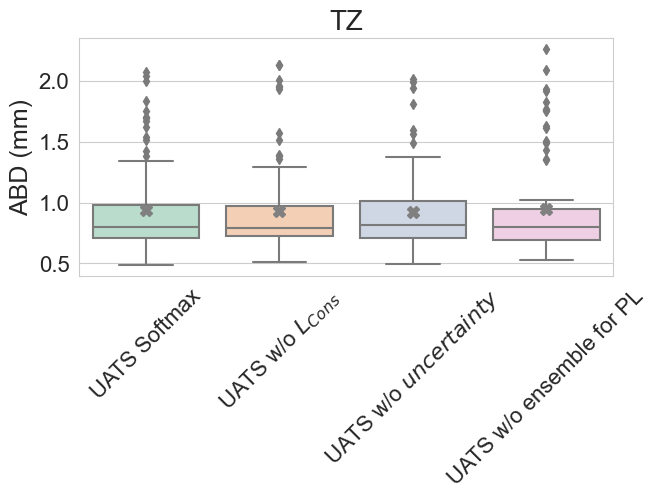}} \\
\subfloat[]{\includegraphics[width = 0.5\textwidth]{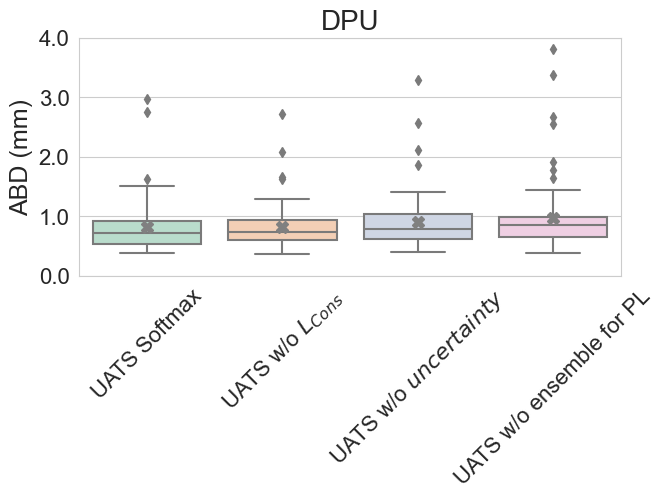}} 
\subfloat[]{\includegraphics[width = 0.5\textwidth]{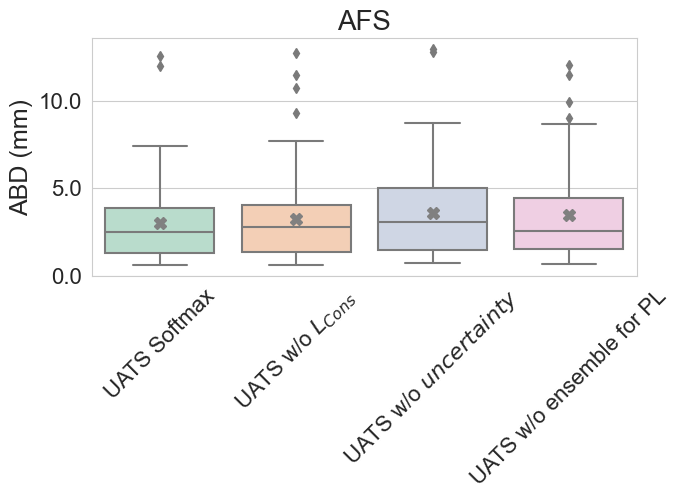}} 

\caption{Boxplots for the segmentation results of the ablation study for UATS. The results are given as the average boundary distance (ABD) in $mm$ between the ground truth and automatic segmentation. The state-of-the-art methods are shown in comparison to our proposed UATS method.} 
\label{fig:boxplots_ablation_abd}
\end{figure}

\end{document}